\documentclass[a4paper,11pt]{article}
\usepackage{pos}
\usepackage[citestyle=numeric-comp,style=numeric-comp,sorting=none,maxbibnames=9,maxcitenames=9,backend=biber]{biblatex}
\begin{document}
\title{\vskip-3cm{\baselineskip14pt
    \begin{flushleft}
     \normalsize P3H-24-038, TTP24-021
    \end{flushleft}} \vskip1.5cm 
    Nonleptonic B-decays at NNLO}
\ShortTitle{Nonleptonic B-decays at NNLO}

\author*[a]{Manuel Egner}
%\author[a,b]{Second Author}

\affiliation[a]{Institut für Theoretische Teilchenphysik, Karlsruhe Institute of Technology (KIT),\\
  Wolfgang-Gaede Straße 1, 76128 Karlsruhe, Germany}

%\affiliation[b]{Department, University,\\
%Street number, City, Country}

\emailAdd{manuel.egner@kit.edu}
%\emailAdd{s.author@univ.country}

\abstract{The decay of B mesons can be predicted within the Heavy Quark Expansion as the decay of a free bottom quark plus corrections which are
suppressed by powers of $1/m_b$. This contribution describes the calculation
of the NNLO QCD corrections to nonleptonic decays of a free bottom quark
including charm quark mass effects. In particular it outlines the
challenges in connection to the computation of master integrals, the
renormalization of the effective operators and the problems
which arise from calculating traces with $\gamma_5$ in $d$ dimensions.}

\FullConference{Loops and Legs in Quantum Field Theory (LL2024)\\
 14-19, April, 2024\\
Wittenberg, Germany\\}

%% \tableofcontents

\maketitle

\section{Introduction}
Lifetimes of $B$ mesons can be calculated in Heavy Quark Effective Expansion (HQE). In this effective theory, the decay width of the $B$ meson, $\Gamma\left(B\right)$, is decomposed into the decay of a free $b$ quark and additional contributions which are suppressed by powers of the heavy quark mass, $m_b$:
\begin{align}
\Gamma\left(B\right) = \Gamma_3 + \Gamma_5\frac{\langle \mathcal{O}_5 \rangle}{m_b^2} + \Gamma_6\frac{\langle \mathcal{O}_6 \rangle}{m_b^3} + \dots  + 16 \pi^2 \left(  \Gamma_6\frac{\langle \tilde{\mathcal{O}}_6 \rangle}{m_b^3} + \Gamma_7\frac{\langle \tilde{\mathcal{O}}_7 \rangle}{m_b^4} +\dots \right).
\end{align}
Since the bottom mass $m_b$ is relatively large compared to the energy scale of the decay, the main contribution to the decay width is $\Gamma_3$, the decay width of the free $b$ quark. In our work, we calculate QCD corrections to this quantity for weak decays of $B$ mesons with a massive charm quark in the final state. These decays can be divided into two different decay channels, the semileptonic and the nonleptonic one. For the semileptonic decay channel, $b \rightarrow c l \bar{\nu}$, QCD corections are known up to $\mathrm{N}^3\mathrm{LO}$ \cite{SLNLO,SLNNLOCP,SLNNLODPC,ourpaper,SLNNNLOFSS,SLNNNLOCCD}.The nonleptonic decays include the two CKM favored decay channels $b \rightarrow c\bar{u}d$ and $b \rightarrow c\bar{c}s$ and CKM suppressed channels, for example $b \rightarrow u\bar{c}s$ and $b \rightarrow u\bar{u}d$. The calculation of these processes is more involved than the semileptonic case. The NLO corrections for $b \rightarrow c\bar{u}d$ and $b \rightarrow c\bar{c}s$ are known \cite{NLcudNLO,NLccsNLO}. At NNLO, first steps were made in Ref. \cite{NLcudNNLO}, however only one effective operator has been considered and massless quarks in the final state have been assumed. 
The uncertainty contributions on B-meson lifetimes are dominated by the uncertainty induced by renormalization scale $\mu$. This uncertainty will be reduced once higher order corrections are known. In the following, the calculation of the NNLO corrections to all nonleptonic decay channels is outlined.
A more detailed discussion can be found in \cite{nleppaper}

\section{Calculation Setup}
The calculation is done by using the optical theorem. This leads to two loop diagrams at LO and therefore four loop diagrams at NNLO. However, only the imaginary part of these diagrams has to be calculated. The diagrams contributing to this process are generated with \texttt{qgraf} \cite{QGRAF}. We find 1308 diagrams at NNLO for each of the decays $b \rightarrow c \overline{u}d$ and $b \rightarrow c \overline{c}s$, which are then mapped to scalar integral families using \texttt{tapir} \cite{tapir} and \texttt{exp}  \cite{exp,exp2}. The diagrams for $b \rightarrow u \overline{c}s$ can be mapped to families of the $b \rightarrow c \overline{u}d$ diagrams and therefore also to the same set of master integrals. The $b \rightarrow u \overline{u}d$ decay channel is obtained by taking the massless limit of one of the other decays and adding an additional contribution originating from closed charm-loop insertions into a gluon propagator. We will call this contribution the $U_c$ contribution in the following.
Using \texttt{Kira} \cite{KIRA1,KIRA2} we find 321 master integrals with non-vanishing imaginary parts for $b \rightarrow c \overline{u}d$, 527 for $b \rightarrow c \overline{c}s$ and 21 for the $U_c$ contribution. Their calculation is described in section \ref{se:masters}.

\section{Evanescent operators and $\gamma_5$}
We describe the nonleptonic decays with the following effective Hamiltonian:
\begin{align}
  \mathcal{H}_{\rm eff} =
  \frac{4 G_F}{\sqrt{2}}
  \sum_{q_{1,3} = u,c}
  \sum_{q_2 = d,s}
  V_{q_1b} V_{q_2 q_3}^*
  \Big(
    C_1(\mu_b) O_1^{q_1 q_2 q_3}
    +C_2(\mu_b) O_2^{q_1 q_2 q_3}
  \Big)
  +{\rm h.c.} \,
\end{align}
with the physical operators
\begin{align}
  O_1^{q_1 q_2 q_3} &=
  (\bar q_1^\alpha \gamma^\mu P_L b^\beta) (\bar q_2^\beta \gamma_\mu P_L q_3^\alpha),
  \notag \\
  O_2^{q_1 q_2 q_3} &=
  (\bar q_1^\alpha \gamma^\mu P_L b^\alpha) (\bar q_2^\beta \gamma_\mu P_L q_3^\beta),
\end{align}
and the matching coefficients $C_i(\mu_b)$. Starting from NLO, the evaluation of diagrams with insertions of these operators includes traces over $\gamma_5$ which have to be evaluated in $d \neq 4$ dimensions. Since the calculation of the anomalous dimension was done using anticommuting $\gamma_5$ \cite{Buras:1989xd,Gorbahn:2004my} we want to apply the same scheme for $\gamma_5$ to be consistent. To avoid the calculation of traces with one $\gamma_5$ we use Fierz identies \cite{NLcudNLO}:
\begin{align}
O_1^{q_1 q_2 q_3} &=
  (\bar q_1^\alpha \gamma^\mu P_L b^\beta) (\bar q_2^\beta \gamma_\mu P_L q_3^\alpha) \xrightarrow{\mathrm{Fierz}} (\bar q_2^\beta \gamma^\mu P_L b^\beta) (\bar q_1^\alpha \gamma_\mu P_L q_3^\alpha) = O_2^{q_2 q_1 q_3}.
\end{align}
After applying this transformation one of the operators in all diagrams, we are left with only one trace over Dirac matrices. In case of two $\gamma_5$ matrices appearing in this trace we can use anticommuting $\gamma_5$. In case of only one $\gamma_5$, we can discard this term, since the decay width we calculate is a parity-even quantity.
Fierz identies are four dimensional but we can restore them order by order in perturbation theory by choosing the correct evanescent operators \cite{Buras:1989xd, Herrlich:1994kh}. In order to do this up to NNLO, we introduce terms proportional to $\epsilon^2$ multiplied with physical operators and undetermined coefficients $\{ A_2, B_1,B_2\}$ to the definition of the evanescent operators:
\begin{align}
E_1^{(1),q_1 q_2 q_3} &=
                          (\bar q_1^\alpha \gamma^{\mu_1 \mu_2 \mu_3} P_L b^\beta)
                          (\bar q_2^\beta
                          \gamma_{\mu_1 \mu_2 \mu_3} P_L
                          q_3^\alpha) - (16 - 4 \epsilon + A_2 \epsilon^2)O_1^{q_1 q_2 q_3},
  \notag \\
  E_2^{(1),q_1 q_2 q_3} &=
                          (\bar q_1^\alpha \gamma^{\mu_1 \mu_2 \mu_3} P_L b^\alpha)
                          (\bar q_2^\beta
                          \gamma_{\mu_1 \mu_2 \mu_3} P_L
                          q_3^\beta) - (16 - 4 \epsilon + A_2 \epsilon^2)O_2^{q_1 q_2 q_3}, \notag\\
  E_1^{(2),q_1 q_2 q_3} &=
                          (\bar q_1^\alpha
                          \gamma^{\mu_1 \mu_2 \mu_3 \mu_4 \mu_5}
                          P_L b^\beta) (\bar q_2^\beta
                          \gamma_{\mu_1 \mu_2\mu_3\mu_4\mu_5}
                          P_L q_3^\alpha) -(256-224 \epsilon + B_1 \epsilon^2 )O_1^{q_1 q_2
                          q_3},  
  \notag \\
  E_2^{(2),q_1 q_2 q_3} &=
                          (\bar q_1^\alpha
                          \gamma^{\mu_1\mu_2\mu_3\mu_4\mu_5}
                          P_L b^\alpha) (\bar q_2^\beta
                          \gamma_{\mu_1\mu_2\mu_3\mu_4\mu_5}
                          P_L q_3^\beta)-(256-224 \epsilon + B_2 \epsilon^2 )O_2^{q_1 q_2 q_3}.
\end{align}
We now fix the coefficients $\{A_2,B_1,B_2\}$ by imposing a symmetric anomalous dimension matrix $\gamma$ \cite{Buras:1989xd,Herrlich:1994kh}
\begin{align}
\mu \frac{\mathrm{d}C_i}{\mathrm{d}\mu} = \gamma_{ij} C_j, & &
\gamma = \left( \begin{array}{cc}
\gamma_{11} & \gamma_{12} \\
\gamma_{21} & \gamma_{22}
\end{array}
\right), \ \ \  \text{with} \
\gamma_{11} = \gamma_{22}, \ \ \gamma_{12} = \gamma_{21}
\end{align}  
This condition ensures the validity of the Fierz symmtry up to NNLO and we obtain
\begin{align}
&A_2=-4, & B_1=-\frac{45936}{125}, & & B_2=-\frac{115056}{115} .
\end{align}
\begin{figure}
\centering
\includegraphics[width=0.8\textwidth]{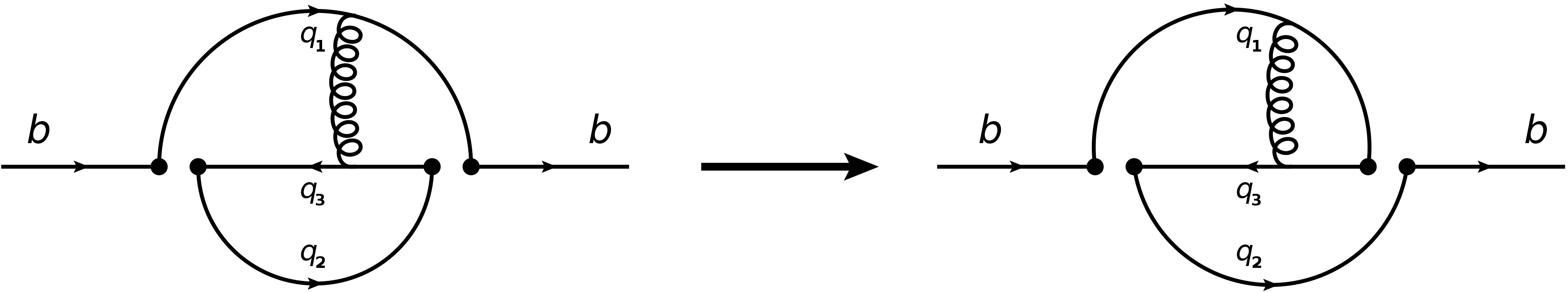}
\caption{Applying Fierz identities to one of the effective operators leads to only one trace.}
\label{fig:Fierz}
\end{figure}

\section{Calculation of master integrals}
\label{se:masters}
We calculate the needed master integrals by using the method developed in Refs. \cite{ExpandMatch,ExpandMatch2}. We construct expansions of the master integrals around different kinematic points using differential equations. To do this, we make an expansion ansatz for the integrals with undetermined coefficients. This ansatz is inserted in the differential equations which yields linear equations between the expansion coefficients. The linear equations can be solved for a small set of independent coefficients using \texttt{Kira} and \texttt{FireFly} \cite{FireFly}.Tey can be determined by matching to precise numerical values of the integrals obtained with \texttt{AMFlow} \cite{AMFlow}.\newline
The ansatz we use for the expansion of the integrals around $\rho=\rho_0$ depends on the expansion point and the singular points of the differential equation. For the various decay channels we have the following singular points:
\begin{itemize}
\item $b \rightarrow c \overline{u}d$: $\rho_{\mathrm{sing}}\in \{0,1/3,1\}$
\item $b \rightarrow c \overline{c}s$: $\rho_{\mathrm{sing}}\in \{0,1/4,1/2\}$
\item $U_c$: $\rho_{\mathrm{sing}} \in \{0,1/2\}$
\end{itemize}
For $\rho \neq \rho_{\mathrm{sing}}$, we can use simple Taylor expansions:
\begin{align}
I_i \left(\rho, \rho_0\right) = \sum_{j=\epsilon_{\mathrm{min}}}^{\epsilon_{\mathrm{max}}}\sum_{m=0}^{j+4}  \sum_{n=0}^{n_{\mathrm{max}}} c \left[ i,j,m,n \right] \epsilon^j \left(\rho_0 -\rho\right)^n,
\end{align}
where the coefficients $c \left[ i,j,m,n \right]$ have both real and imaginary parts. 
For the expansion around the three-charm threshhold corresponding to the singular point at $\rho=m_c/m_b=1/3$ in the $b \rightarrow c \overline{u}d$ decay channel, we use the ansatz
\begin{align}
I_i \left(\rho, \rho_0\right) = \sum_{j=\epsilon_{\mathrm{min}}}^{\epsilon_{\mathrm{max}}}\sum_{m=0}^{j+4}  \sum_{n=0}^{n_{\mathrm{max}}} c \left[ i,j,m,n \right] \epsilon^j \left(\rho -\rho_0\right)^n\log^m\left(\rho-\rho_0\right),
\label{AnsatzPoly}
\end{align}
with $\rho_0=1/3$. When crossing the three-charm threshold at $\rho=1/3$ from $\rho>1/3$ to $\rho<1/3$, the argument of the logarithm gets negative and produces an additional imaginary part, which corresponds to the three-charm contribution. \newline
For the expansions around $\rho=0$, we use the same ansatz as given in equation \eqref{AnsatzPoly} with $\rho_0=0$. \newline
For expansions around a threshold with an even number of massive particles in the final state, we need to include roots in our ansatz:
\begin{align}
I_i \left(\rho, \rho_0\right) = \sum_{j=\epsilon_{\mathrm{min}}}^{\epsilon_{\mathrm{max}}}\sum_{m=0}^{j+4}  \sum_{n=0}^{n_{\mathrm{max}}} c \left[ i,j,m,n \right] \epsilon^j \left(\sqrt{\rho -\rho_0}\right)^n\log^m\left(\rho-\rho_0\right),
\label{WurzelAnsatz}
\end{align}
In our calculation we construct expansions for the master integrals around the following points:
\begin{itemize}
\item $b \rightarrow c \overline{u}d$: $\rho_{0}\in \{0,1/4,1/3,1/2,7/10,1\}$
\item $b \rightarrow c \overline{c}s$: $\rho_{0}\in \{0,1/5,1/3\}$
\item $U_c$: $\rho_{0}\in \{0,1/3,1/2,7/10,1\}$
\end{itemize}

\section{Results}
\begin{figure}
\centering
\includegraphics[width=0.7\textwidth]{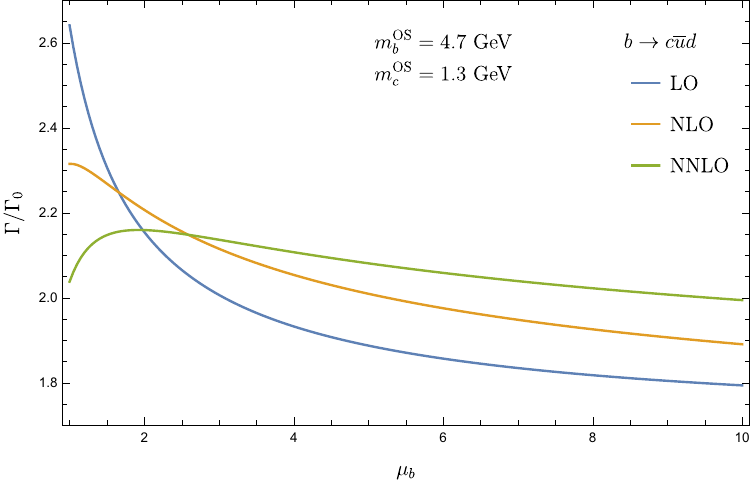}
\caption{The decay width for the channel $b \rightarrow c \overline{u}d$ at LO, NLO and NNLO as function of the renormalization scale $\mu_b$. The mass ratio is set to $\rho=m_c/m_b=1.3/4.7$. This figure is taken from \cite{nleppaper}.}
\label{fig:Decaywidth}
\end{figure}

For simplicity we only show the numerically most important decay channel $b \rightarrow c \bar{u} d$ here.
The result for the decay width can be written in the form
\begin{align*}
\Gamma \left(b \rightarrow c \bar{u} d \right) &= \frac{G_f^2 m_b^5 \left| V_{bc}\right|^2}{192 \pi^3} \left[C_1^2(\mu) G_{11} + C_1(\mu)C_2(\mu) G_{12} +C_2^2(\mu) G_{22}\right].
\end{align*} 
To get an estimate of the corrections, we set the on-shell masses of the quarks to $m_c=1.3 \mathrm{GeV}$ and $m_b=4.7 \mathrm{GeV}$. Evaluating the decay width at $\mu=m_b$, we obtain
\begin{align}
\Gamma \left(b \rightarrow c \bar{u} d \right)  = \Gamma_0 \left[ 1.89907 + 1.77538 \left( \frac{\alpha_s}{\pi} \right) +14.1081 \left( \frac{\alpha_s}{\pi} \right)^2 \right]\Bigg|_{\mu=m_b} ,
\end{align}
where $\Gamma_0 = G_F^2m_b^5 \left|V_{cb}\right|^2 \left|V_{ud}\right|^2 /(192 \pi^3)$. In Figure \ref{fig:Decaywidth} we show the decay width as a function of the renormalization scale $\mu_b$. To estimate the uncertainty induced by the renormalization scale, we consider the region $m_b/2 < \mu_b< 2m_b$. One observes a relative uncertainty of $\approx 7 \%$ at LO which reduces to $\approx 3.5\%$ at NNLO relative to the central value $\mu=m_b$.

\section*{Acknowledgements}
This work was done in collaboration with Matteo Fael, Kay Schönwald and Matthias Steinhauser. This research was supported by the Deutsche
Forschungsgemeinschaft (DFG, German Research Foundation) under grant 396021762 —
TRR 257 “Particle Physics Phenomenology after the Higgs Discovery”. 

\newpage
\printbibliography[heading=bibintoc]

@article{SLNLO,
    author = "Nir, Yosef",
    title = "{The Mass Ratio m(c) / m(b) in Semileptonic B Decays}",
    reportNumber = "SLAC-PUB-4847",
    doi = "10.1016/0370-2693(89)91495-0",
    journal = "Phys. Lett. B",
    volume = "221",
    pages = "184--190",
    year = "1989"
}

@article{QGRAF,
title = {Automatic Feynman Graph Generation},
journal = {Journal of Computational Physics},
volume = {105},
number = {2},
pages = {279-289},
year = {1993},
issn = {0021-9991},
doi = {https://doi.org/10.1006/jcph.1993.1074},
url = {https://www.sciencedirect.com/science/article/pii/S0021999183710740},
author = {P. Nogueira}
}

@inproceedings{exp,
    author = "Seidensticker, T.",
    title = "{Automatic application of successive asymptotic expansions of Feynman diagrams}",
    booktitle = "{6th International Workshop on New Computing Techniques in Physics Research: Software Engineering, Artificial Intelligence Neural Nets, Genetic Algorithms, Symbolic Algebra, Automatic Calculation}",
    eprint = "hep-ph/9905298",
    archivePrefix = "arXiv",
    reportNumber = "TTP-99-22",
    month = "5",
    year = "1999"
}

@article{exp2,
    author = "Harlander, R. and Seidensticker, T. and Steinhauser, M.",
    title = "{Complete corrections of Order alpha alpha-s to the decay of the Z boson into bottom quarks}",
    eprint = "hep-ph/9712228",
    archivePrefix = "arXiv",
    reportNumber = "MPI-PHT-97-81, TTP-97-52",
    doi = "10.1016/S0370-2693(98)00220-2",
    journal = "Phys. Lett. B",
    volume = "426",
    pages = "125--132",
    year = "1998"
}

@article{tapir,
	doi = {10.1016/j.cpc.2022.108544},
  
	url = {https://doi.org/10.1016%2Fj.cpc.2022.108544},
  
	year = 2023,
	month = {jan},
  
	publisher = {Elsevier {BV}},
  
	volume = {282},
  
	pages = {108544},
  
	author = {Marvin Gerlach and Florian Herren and Martin Lang},
  
	title = {tapir: A tool for topologies, amplitudes, partial fraction decomposition and input for reductions},
  
	journal = {Computer Physics Communications}}

@article{KIRA1,
   title={Kira—A Feynman integral reduction program},
   volume={230},
   ISSN={0010-4655},
   url={http://dx.doi.org/10.1016/j.cpc.2018.04.012},
   DOI={10.1016/j.cpc.2018.04.012},
   journal={Computer Physics Communications},
   publisher={Elsevier BV},
   author={Maierhöfer, P. and Usovitsch, J. and Uwer, P.},
   year={2018},
   month={Sep},
   pages={99–112}
}

@article{KIRA2,
   title={Integral reduction with Kira 2.0 and finite field methods},
   volume={266},
   ISSN={0010-4655},
   url={http://dx.doi.org/10.1016/j.cpc.2021.108024},
   DOI={10.1016/j.cpc.2021.108024},
   journal={Computer Physics Communications},
   publisher={Elsevier BV},
   author={Klappert, J. and Lange, F. and Maierhöfer, P. and Usovitsch, J.},
   year={2021},
   month={Sep},
   pages={108024}
}

@article{AMFlow,
	doi = {10.1016/j.cpc.2022.108565},
  
	url = {https://doi.org/10.1016%2Fj.cpc.2022.108565},
  
	year = 2023,
	month = {feb},
  
	publisher = {Elsevier {BV}},
  
	volume = {283},
  
	pages = {108565},
  
	author = {Xiao Liu and Yan-Qing Ma},
  
	title = {{AMFlow}: A Mathematica package for Feynman integrals computation via auxiliary mass flow},
  
	journal = {Computer Physics Communications}
	}

@article{SLNNLOCP,
	doi = {10.1103/physrevlett.100.241807},
  
	url = {https://doi.org/10.1103%2Fphysrevlett.100.241807},
  
	year = 2008,
	month = {jun},
  
	publisher = {American Physical Society ({APS})},
  
	volume = {100},
  
	number = {24},
  
	author = {Alexey Pak and Andrzej Czarnecki},
  
	title = {Mass Effects in Muon and Semileptonic $b \rightarrow c$ Decays},
  
	journal = {Physical Review Letters}
	}

@article{SLNNLODPC,
	doi = {10.1103/physrevd.78.074024},
  
	url = {https://doi.org/10.1103%2Fphysrevd.78.074024},
  
	year = 2008,
	month = {oct},
  
	publisher = {American Physical Society ({APS})},
  
	volume = {78},
  
	number = {7},
  
	author = {Matthew Dowling and Jan H. Piclum and Andrzej Czarnecki},
  
	title = {Semileptonic decays in the limit of a heavy daughter quark},
  
	journal = {Physical Review D}
	}

@article{SLNNNLOFSS,
  title = {Third order corrections to the semileptonic $b\ensuremath{\rightarrow}c$ and the muon decays},
  author = {Fael, Matteo and Sch\"onwald, Kay and Steinhauser, Matthias},
  journal = {Phys. Rev. D},
  volume = {104},
  issue = {1},
  pages = {016003},
  numpages = {7},
  year = {2021},
  month = {Jul},
  publisher = {American Physical Society},
  doi = {10.1103/PhysRevD.104.016003},
  url = {https://link.aps.org/doi/10.1103/PhysRevD.104.016003}
}

@article{SLNNNLOCCD,
	doi = {10.1103/physrevd.103.l111301},
  
	url = {https://doi.org/10.1103%2Fphysrevd.103.l111301},
  
	year = 2021,
	month = {jun},
  
	publisher = {American Physical Society ({APS})},
  
	volume = {103},
  
	number = {11},
  
	author = {Micha{\l} Czakon and Andrzej Czarnecki and Matthew Dowling},
  
	title = {Three-loop corrections to the muon and heavy quark decay rates},
  
	journal = {Physical Review D}
}

@article{NLccsNLO,
	doi = {10.1016/0370-2693(95)00437-p},
  
	url = {https://doi.org/10.1016%2F0370-2693%2895%2900437-p},
  
	year = 1995,
	month = {jun},
  
	publisher = {Elsevier {BV}},
  
	volume = {351},
  
	number = {4},
  
	pages = {546--554},
  
	author = {E. Bagan and Patricia Ball and B. Fiol and P. Gosdzinsky},
  
	title = {Next-to-leading order radiative corrections to the decay b $\rightarrow$ ccs},
  
	journal = {Physics Letters B}
}

@article{NLcudNLO,
	doi = {10.1016/0550-3213(94)90591-6},
  
	url = {https://doi.org/10.1016%2F0550-3213%2894%2990591-6},
  
	year = 1994,
	month = {dec},
  
	publisher = {Elsevier {BV}},
  
	volume = {432},
  
	number = {1-2},
  
	pages = {3--38},
  
	author = {E. Bagan and Patricia Ball and V.M. Braun and P. Gosdzinsky},
  
	title = {Charm quark mass dependence of QCD corrections to nonleptonic inclusive B decays},
  
	journal = {Nuclear Physics B}
}

@article{NLcudNNLO,
	doi = {10.1103/physrevlett.96.171803},
  
	url = {https://doi.org/10.1103%2Fphysrevlett.96.171803},
  
	year = 2006,
	month = {may},
  
	publisher = {American Physical Society ({APS})},
  
	volume = {96},
  
	number = {17},
  
	author = {Andrzej Czarnecki and Maciej {\'{S}}lusarczyk and Fyodor Tkachov},
  
	title = {Enhancement of hadronic b quark Decays},
  
	journal = {Physical Review Letters}
}

@article{ExpandMatch,
	doi = {10.1007/jhep09(2021)152},
  
	url = {https://doi.org/10.1007%2Fjhep09%282021%29152},
  
	year = 2021,
	month = {sep},
  
	publisher = {Springer Science and Business Media {LLC}},
  
	volume = {2021},
  
	number = {9},
  
	author = {Matteo Fael and Fabian Lange and Kay Schönwald and Matthias Steinhauser},
  
	title = {A semi-analytic method to compute Feynman integrals applied to four-loop corrections to the $\overline{\mathrm{MS}}$-pole quark mass relation},
  
	journal = {Journal of High Energy Physics}
}

@article{ExpandMatch2,
    author = {Fael, Matteo and Lange, Fabian and Sch\"onwald, Kay and Steinhauser, Matthias},
    title = "{Singlet and nonsinglet three-loop massive form factors}",
    eprint = "2207.00027",
    archivePrefix = "arXiv",
    primaryClass = "hep-ph",
    reportNumber = "TTP22-042, P3H-22-066",
    doi = "10.1103/PhysRevD.106.034029",
    journal = "Phys. Rev. D",
    volume = "106",
    number = "3",
    pages = "034029",
    year = "2022"
}

@article{FireFly,
	doi = {10.1016/j.cpc.2021.107968},
  
	url = {https://doi.org/10.1016%2Fj.cpc.2021.107968},
  
	year = 2021,
	month = {jul},
  
	publisher = {Elsevier {BV}},
  
	volume = {264},
  
	pages = {107968},
  
	author = {Jonas Klappert and Sven Yannick Klein and Fabian Lange},
  
	title = {Interpolation of dense and sparse rational functions and other improvements in {FireFly}},
  
	journal = {Computer Physics Communications}
}

@article{ourpaper,
    author = {Egner, Manuel and Fael, Matteo and Sch\"onwald, Kay and Steinhauser, Matthias},
    note="accepted for publication in Journal of High Energy Physics",
    title = {Revisiting semileptonic $B$ meson decays at next-to-next-to-leading order},
    eprint = "2308.01346",
    archivePrefix = "arXiv",
    primaryClass = "hep-ph",
    reportNumber = "P3H-23-053, TTP23-030, ZU-TH 41/23, CERN-TH-2023-151",
    month = "8",
    year = "2023"
}

@article{Gorbahn:2004my,
    author = "Gorbahn, Martin and Haisch, Ulrich",
    title = "{Effective Hamiltonian for non-leptonic $|\Delta F| = 1$ decays at NNLO in QCD}",
    eprint = "hep-ph/0411071",
    archivePrefix = "arXiv",
    reportNumber = "IPPP-04-66, DCPT-04-132, FERMILAB-PUB-04-281-T",
    doi = "10.1016/j.nuclphysb.2005.01.047",
    journal = "Nucl. Phys. B",
    volume = "713",
    pages = "291--332",
    year = "2005"
}

@article{Buras:1989xd,
    author = "Buras, Andrzej J. and Weisz, Peter H.",
    title = "{QCD Nonleading Corrections to Weak Decays in Dimensional Regularization and 't Hooft-Veltman Schemes}",
    reportNumber = "MPI-PAE/PTh-42/89, TUM-T32-189",
    doi = "10.1016/0550-3213(90)90223-Z",
    journal = "Nucl. Phys. B",
    volume = "333",
    pages = "66--99",
    year = "1990"
}

@article{Herrlich:1994kh,
    author = "Herrlich, Stefan and Nierste, Ulrich",
    title = "{Evanescent operators, scheme dependences and double insertions}",
    eprint = "hep-ph/9412375",
    archivePrefix = "arXiv",
    reportNumber = "TUM-T31-66-94, PSI-PR-94-37",
    doi = "10.1016/0550-3213(95)00474-7",
    journal = "Nucl. Phys. B",
    volume = "455",
    pages = "39--58",
    year = "1995"
}

@misc{nleppaper,
      title={Nonleptonic $B$-meson decays to next-to-next-to-leading order}, 
      author={Manuel Egner and Matteo Fael and Kay Schönwald and Matthias Steinhauser},
      year={2024},
      eprint={2406.19456},
      archivePrefix={arXiv},
      primaryClass={hep-ph},
      url={https://arxiv.org/abs/2406.19456}, 
}

\end{document}